\documentclass{amsart}
\usepackage{amssymb}
\usepackage{amsmath}
\usepackage{amsfonts}

\newtheorem{theorem}{Theorem}
\theoremstyle{plain}

\newtheorem{definition}[theorem]{Definition}
\newtheorem{example}[theorem]{Example}

\newtheorem{lemma}[theorem]{Lemma}

\newtheorem{remark}[theorem]{Remark}

\numberwithin{equation}{section}
\numberwithin{theorem}{section}
\input{tcilatex}

\begin{document}
\title{On some deterministic dictionaries supporting sparsity}
\author{Shamgar Gurevich}
\address{Department of Mathematics, University of California, Berkeley, CA
94720, USA. }
\email{shamgar@math.berkeley.edu}
\author{Ronny Hadani}
\address{Department of Mathematics, University of Chicago, IL 60637, USA.}
\email{hadani@math.uchicago.edu}
\author{Nir Sochen}
\address{School of Mathematical Sciences, Tel Aviv University, Tel Aviv
69978, Israel.}
\email{sochen@math.tau.ac.il}
\dedicatory{To appear in the special issue of JFAA on sparsity}
\date{September 1, 2007}
\keywords{Sparsity, Deterministic dictionaries, Low coherence, Weil
representation, Commutative subgroups, Eigenfunctions, Explicit algorithm. }
\thanks{Mathematics Subject Classification (2000) 94A12, 11F27.}
\thanks{\copyright \ Copyright by S. Gurevich, R. Hadani and N. Sochen,
September 1, 2007. All rights reserved.}

\begin{abstract}
We describe a new construction of an incoherent dictionary, referred to as
the \textit{oscillator dictionary}, which is based on considerations in the
representation theory of finite groups. The oscillator dictionary consists
of approximately $p^{5}$ unit vectors in a Hilbert space of dimension $p$,
whose pairwise inner products have magnitude of at most $4/\sqrt{p}$. An
explicit algorithm to construct a large portion of the oscillator dictionary
is presented.
\end{abstract}

\maketitle

\section{Introduction}

Digital signals, or simply signals, can be thought of as functions on the
finite line $\mathbb{F}_{p}$, namely the finite field with $p$ elements,
where $p$ is a prime number. The space of signals $\mathcal{H=%
\mathbb{C}
}\left( \mathbb{F}_{p}\right) $ is a Hilbert space, with the inner product
given by the standard formula 
\begin{equation*}
\left \langle f,g\right \rangle =\tsum \limits_{t\in \mathbb{F}_{p}}f\left(
t\right) \overline{g\left( t\right) }.
\end{equation*}

\subsection{Incoherent dictionaries}

A central problem is to construct useful classes of signals that demonstrate
strong descriptive power and at the same time are characterized by formal
mathematical conditions. Meeting these two requirements is a non trivial
task and is a source for many novel developments in the field of signal
processing. The problem was tackled, over the years, by various approaches.

Two decades ago \cite{DGM}, a novel approach was introduced, hinting towards
a fundamental change of perspective about the nature of signals. In this
approach, a signal is characterized in terms of its sparsest presentation as
a linear combination of vectors (also called \textit{atoms}) in a \textit{%
dictionary}. The characterization is intrinsically non-linear, hence, as a
consequence, one comes to deal with classes of signals which are not closed
with respect to addition. More formally:

\begin{definition}
A set of vectors $\mathfrak{D\subset }\mathcal{H}$ is called an $N$-\textbf{%
independent} dictionary if every subset $\mathfrak{D}^{\prime }\subset 
\mathfrak{D}$, with $\left \vert \mathfrak{D}^{\prime }\right \vert =N$, is
linearly independent.
\end{definition}

This notion is very close to the notion of \textit{spark} of a dictionary
introduced in \cite{DE}.

Given an $2N$-independent dictionary $\mathfrak{D}$, every signal $f\in 
\mathcal{H}$, has at most one presentation of the form 
\begin{equation*}
f=\tsum \limits_{\varphi \in \mathfrak{D}^{\prime }}a_{\varphi }\varphi ,
\end{equation*}%
for $\mathfrak{D}^{\prime }\subset \mathfrak{D}$ with $\left \vert \mathfrak{D%
}^{\prime }\right \vert \leq N$. Such a presentation, if exists, is unique
and is called the \textbf{sparse} presentation. Consequently, we will also
call such a dictionary $\mathfrak{D}$ an $N$-\textbf{sparse} dictionary.
Given that a signal $f$ admits a sparse presentation, a basic difficulty is
to effectively reconstruct the sparse coefficients $a_{\varphi }$. A way to
overcome this difficulty is to introduce \cite{BDE, DE, EB, GMS, GG, GN, Tr}
the stronger notion of \textit{incoherent dictionary}.

\begin{definition}
A set of vectors $\mathfrak{D\subset }\mathcal{H}$ is called $\mu $-\textbf{%
coherent }dictionary\textbf{, }for $0\leq \mu \ll 1,$ if for every two
different vectors $\varphi ,\phi \in \mathfrak{D}$ we have $\left \vert
\left \langle \varphi ,\phi \right \rangle \right \vert \leq \mu $.
\end{definition}

The two notions of coherence and sparsity are related by the following
proposition \cite{BDE, DE, EB, GN, Tr}

\begin{theorem}
\label{eff_prop}If $\mathfrak{D}$ is $1/R$-coherent then $\mathfrak{D}$ is $%
\left \lfloor R/2\right \rfloor $-sparse\footnote{%
Here $\left \lfloor R/2\right \rfloor $ stands for the greatest integer
which is less then or equal to $R/2$.}, moreover there exists an effective
algorithm to extract the sparse coefficients.
\end{theorem}

A basic problem \cite{BDE, DGM, S, SH} in the theory is introducing
systematic constructions of "good" incoherent dictionaries. Here "good"
means that the size of the dictionary and the sparsity factor $N$ are made
as large as possible.

In this paper, we begin to develop a systematic approach to the construction
of incoherent dictionaries based on the representation theory of groups over
finite fields. In particular, we describe an examples of such dictionary
called the oscillator dictionary.

\subsection{Main results}

The main contribution of this paper is the introduction of a dictionary $%
\mathfrak{D}_{O}$, that we call the \textit{oscillator }dictionary, which is
constructed using the representation theory of the two dimensional
symplectic group $SL_{2}\left( \mathbb{F}_{p}\right) $. The oscillator
dictionary is $4/\sqrt{p}$-coherent, consisting of approximately $p^{3}$
vectors. We also introduce an extended oscillator dictionary $\mathfrak{D}%
_{E}$ which is $4/\sqrt{p}$-coherent and consists of approximately $p^{5}$
vectors. Our goal is to explain the construction of $\mathfrak{D}_{O}$ and
state some of its properties which are relevant to sparsity, referring the
reader to \cite{GHS} for a more comprehensive treatment.

As a suggestive model example we explain first the construction of the well
known \textit{Heisenberg }dictionary $\mathfrak{D}_{H}$ (see \cite{H, HCM}),
which is constructed using the representation theory of the finite \textit{%
Heisenberg group }over the finite field $\mathbb{F}_{p}$.\textit{\ }The
Heisenberg dictionary is\textit{\ }$1/\sqrt{p}$-coherent, consisting of
approximately $p^{2}$ vectors.

\subsection{\textbf{Structure of the paper }}

The paper consists of two sections and two appendices. In Section \ref{H-W},
several basic notions from representation theory are introduced.
Particularly, we present the Heisenberg and Weil representations over finite
fields. In Section \ref{H-O-D}, we introduce the Heisenberg and oscillator
dictionaries $\mathfrak{D}_{H}$ and $\mathfrak{D}_{O}$ respectively, and the
extended dictionary $\mathfrak{D}_{E}.$ In Appendix \ref{TGR_sec}, we
explain in more details basic concepts from group representation theory that
we use in the body of the paper. Finally, in Appendix \ref{Algorithm},\ we
describe an explicit\ algorithm that generates a large portion of the
oscillator dictionary.

\begin{remark}[Field extension]
\textbf{\ }All the results in this paper were stated for the basic finite
field $\mathbb{F}_{p},$ for the reason of making the terminology more
accessible. However, they are valid \cite{GHS} for any field extension of
the form $\mathbb{F}_{q}$ with $q=p^{n}.$ One should only replace $p$ by $q$
in all appropriate places.
\end{remark}

{\Large Acknowledgement.}{\LARGE \ }It is a pleasure to thank J. Bernstein
for his interest and guidance in the mathematical aspects of this work. We
are grateful to S. Golomb and G. Gong for their interest in this project. We
would like to thank M. Elad, O. Holtz, R. Kimmel, L.H. Lim, and A. Sahai for
interesting discussions. Finally, we thank B. Sturmfels for encouraging us
to proceed in this line of research.

\section{The Heisenberg and Weil representations\label{H-W}}

\subsection{The Heisenberg group}

Let $(V,\omega )$ be a two-dimensional symplectic vector space over the
finite field $\mathbb{F}_{p}$. The reader should think of $V$ as $\mathbb{F}%
_{p}\times \mathbb{F}_{p}$ with the standard symplectic form 
\begin{equation*}
\omega \left( \left( \tau ,w\right) ,\left( \tau ^{\prime },w^{\prime
}\right) \right) =\tau w^{\prime }-w\tau ^{\prime }.
\end{equation*}

Considering $V$ as an Abelian group, it admits a non-trivial central
extension called the \textit{Heisenberg }group. \ Concretely, the group $H$
can be presented as the set $H=V\times \mathbb{F}_{p}$ with the
multiplication given by%
\begin{equation*}
(v,z)\cdot (v^{\prime },z^{\prime })=(v+v^{\prime },z+z^{\prime }+\tfrac{1}{2%
}\omega (v,v^{\prime })).
\end{equation*}

The center of $H$ is $\ Z=Z(H)=\left \{ (0,z):\text{ }z\in \mathbb{F}%
_{p}\right \} .$ The symplectic group $Sp=Sp(V,\omega )$, which in this case
is just isomorphic to $SL_{2}\left( \mathbb{F}_{p}\right) $, acts by
automorphism of $H$ through its action on the $V$-coordinate, that is, a
matrix 
\begin{equation*}
g=%
\begin{pmatrix}
a & b \\ 
c & d%
\end{pmatrix}%
,
\end{equation*}%
sends an element $\left( v,z\right) $, where $v=\left( \tau ,w\right) $ to
the element $\left( gv,z\right) $, where $gv=\left( a\tau +bw,c\tau
+dw\right) $.

\subsection{The Heisenberg representation\label{HR}}

One of the most important attributes of the group $H$ is that it admits,
principally, a unique irreducible representation (see Subsection \ref%
{rep_sub}). The precise statement goes as follows: Let $\psi :Z\rightarrow
S^{1},$ where $S^{1}$ denotes the unit circle$,$ be a non-trivial unitary
character of the center, that is $\psi \neq 1$ and satisfies $\psi \left(
z_{1}+z_{2}\right) =\psi \left( z_{1}\right) \cdot \psi \left( z_{2}\right) $%
, for every $z_{1},z_{2}\in Z$; for example, in this paper we take $\psi
\left( z\right) =e^{\frac{2\pi i}{p}z}$.

We denote by $U\left( \mathcal{H}\right) $ the group of unitary operators on 
$\mathcal{H}$. It is not difficult to show \cite{T} that

\begin{theorem}[Stone-von Neuman]
\label{S-vN}There exists a unique (up to isomorphism) irreducible unitary
representation $\pi :H\rightarrow U\left( \mathcal{H}\right) $ with central
character $\psi $, that is, $\pi \left( z\right) =\psi \left( z\right) \cdot
Id_{\mathcal{H}}$, for every $z\in Z$.
\end{theorem}

The representation $\pi $ which appears in the above theorem will be called
the \textit{Heisenberg representation}.

More concretely, $\pi :H\rightarrow U\left( \mathcal{H}\right) $ can be
realized as follows: $\mathcal{H}$ is the Hilbert space $%
\mathbb{C}
(\mathbb{F}_{p})$ of complex valued functions on the finite line, with the
standard inner product%
\begin{equation*}
\left \langle f,g\right \rangle =\sum \limits_{t\in \mathbb{F}_{p}}f\left(
t\right) \overline{g\left( t\right) }\text{,}
\end{equation*}%
for every $f,g\in 
\mathbb{C}
(\mathbb{F}_{p})$ and the action $\pi $ is given by

\begin{itemize}
\item $\pi (\tau ,0)[f]\left( t\right) =f\left( t+\tau \right) ;$

\item $\pi (0,w)[f]\left( t\right) =\psi \left( wt\right) f\left( t\right) ;$

\item $\pi (z)[f]\left( t\right) =\psi \left( z\right) f\left( t\right) ,$ $%
z\in Z.$
\end{itemize}

Here we are using $\tau $ to indicate the first coordinate and $w$ to
indicate the second coordinate of $\ V\simeq \mathbb{F}_{p}\times \mathbb{F}%
_{p}$.

We will call this explicit realization the \textit{standard realization}.

\subsection{The Weil representation\label{Wrep_sub}}

A direct consequence of Theorem \ref{S-vN} is the existence of a projective
unitary representation $\widetilde{\rho }:Sp\rightarrow U(\mathcal{H)}$,
that is, a collection of operators $\left \{ \widetilde{\rho }(g)\in U\left( 
\mathcal{H}\right) :g\in Sp\right \} $ which satisfy multiplicativity up-to
a unitary scalar 
\begin{equation*}
\widetilde{\rho }(gh)=C\left( g,h\right) \cdot \widetilde{\rho }(g)\circ 
\widetilde{\rho }(h),
\end{equation*}%
for every $g,h\in Sp$ and $C\left( g,h\right) \in S^{1}$. The construction
of $\widetilde{\rho }$ out of the Heisenberg representation $\pi $ is due to
Weil \cite{W} and it goes as follows: Considering the Heisenberg
representation $\pi :H\rightarrow U\left( \mathcal{H}\right) $ and an
element $g\in Sp$, one can define a new representation $\pi
^{g}:H\rightarrow U\left( \mathcal{H}\right) $ by $\pi ^{g}\left( h\right)
=\pi \left( g\left( h\right) \right) $. Clearly both $\pi $ and $\pi ^{g}$
have the same central character $\psi $ hence, by Theorem \ref{S-vN}, they
are isomorphic. Since the space of intertwining morphisms (see Subsection %
\ref{inter_sub}) $\mathsf{Hom}_{H}(\pi ,\pi ^{g})$ is one dimensional (this
follows from Schur's lemma, see Subsection \ref{rep_sub}), choosing for
every $g\in Sp$ a non-zero representative $\widetilde{\rho }(g)\in \mathsf{%
Hom}_{H}(\pi ,\pi ^{g})$ gives the required projective representation.

In more concrete terms, the projective representation $\widetilde{\rho }$ is
characterized by the formula 
\begin{equation}
\widetilde{\rho }\left( g\right) \pi \left( h\right) \widetilde{\rho }\left(
g^{-1}\right) =\pi \left( g\left( h\right) \right) ,  \label{Egorov}
\end{equation}%
for every $g\in Sp$ and $h\in H$. \ 

The important and non-trivial statement is that the projective
representation $\widetilde{\rho }$ \ can be linearized in a unique manner
into an honest unitary representation:

\begin{theorem}
\label{linearization} There exists a unique\footnote{%
Unique, except in the case the finite field is $\mathbb{F}_{3}$.} unitary
representation 
\begin{equation*}
\rho :Sp\longrightarrow U(\mathcal{H)},
\end{equation*}

such that every operator $\rho \left( g\right) $ satisfies Equation (\ref%
{Egorov}).
\end{theorem}

For the sake of concreteness, let us give an explicit description of the
operators $\rho \left( g\right) $, for different elements $g\in Sp$, as they
appear in the standard realization. The operators will be specified up to a
unitary scalar.

\begin{itemize}
\item The standard diagonal subgroup $A\subset Sp$ acts by (normalized)
scaling: An element 
\begin{equation*}
\begin{pmatrix}
{\small a} & {\small 0} \\ 
{\small 0} & {\small a}^{-1}%
\end{pmatrix}%
,
\end{equation*}%
acts by 
\begin{equation*}
S_{a}\left[ f\right] \left( t\right) =\sigma \left( a\right) f\left(
a^{-1}t\right) ,
\end{equation*}%
where $\sigma :\mathbb{F}_{p}^{\times }\rightarrow \{ \pm 1\}$ is the unique
non-trivial quadratic character of the multiplicative group $\mathbb{F}%
_{p}^{\times }$ (also called the Legendre character), given by $\sigma (a)=$ 
$a^{\frac{p-1}{2}}(\func{mod}$ $p)$.

\item The subgroup of strictly lower diagonal elements $U\subset Sp$ acts by
quadratic exponents (chirps): An element 
\begin{equation*}
u=%
\begin{pmatrix}
1 & 0 \\ 
u & 1%
\end{pmatrix}%
,
\end{equation*}%
acts by 
\begin{equation*}
M_{u}\left[ f\right] \left( t\right) =\psi (-\tfrac{u}{2}t^{2})f\left(
t\right) .
\end{equation*}

\item The Weyl element 
\begin{equation*}
\mathrm{w}=%
\begin{pmatrix}
0 & 1 \\ 
-1 & 0%
\end{pmatrix}%
,
\end{equation*}%
acts by discrete Fourier transform 
\begin{equation*}
F\left[ f\right] \left( w\right) =\frac{1}{\sqrt{p}}\sum \limits_{t\in 
\mathbb{F}_{p}}\psi \left( wt\right) f\left( t\right) .
\end{equation*}
\end{itemize}

\section{The Heisenberg and the oscillator dictionaries\label{H-O-D}}

\subsection{Model example: The Heisenberg dictionary}

The Heisenberg dictionary is a collection of $p+1$ orthonormal bases, each
characterized, roughly, as eigenvectors of a specific linear operator. An
elegant way to define this dictionary is using the Heisenberg representation 
\cite{H, HCM}.

\subsubsection{Bases associated with lines}

The Heisenberg group is non-commutative, yet it consists of various
commutative subgroups which can be easily described as follows: Let $%
L\subset V$ be a line through the origin in $V$. One can associate to $L$ a
commutative subgroup $A_{L}\subset H$, given by $A_{L}=\left \{ \left(
l,0\right) :l\in L\right \} $. It will be convenient to identify the group $%
A_{L}$ with the line $L$. Restricting the Heisenberg representation $\pi $
to the commutative subgroup $L$, namely, considering the restricted
representation $\pi :L\rightarrow U\left( \mathcal{H}\right) $, one obtains
a collection of pairwise commuting operators $\left \{ \pi \left( l\right)
:l\in L\right \} $, which, in turns, yields an orthogonal decomposition into
character spaces (see Subsection \ref{char_sub}) 
\begin{equation*}
\mathcal{H=}\tbigoplus \limits_{\chi }\mathcal{H}_{\chi },
\end{equation*}%
where $\chi $ runs in the set $\widehat{L}$ of unitary characters of $L$,
that is, each $\chi \in $ $\widehat{L}$ is a function $\chi :L\rightarrow
S^{1}$ which satisfies $\chi \left( l_{1}+l_{2}\right) =\chi \left(
l_{1}\right) \cdot \chi \left( l_{2}\right) $, for every $l_{1},l_{2}\in L$.

A more concrete way to specify the above decomposition is by choosing a
non-zero vector $l_{0}\in L$. After such a choice, the character space $%
\mathcal{H}_{\chi }$ naturally corresponds to the eigenspace of the linear
operator $\pi \left( l_{0}\right) $ associated with the eigenvalue $\lambda
=\chi \left( l_{0}\right) $.

It is not difficult to verify in this case that

\begin{lemma}
For every $\chi \in \widehat{L}$ we have $\dim \mathcal{H}_{\chi }=1$.
\end{lemma}

Choosing a vector $\varphi _{\chi }\in \mathcal{H}_{\chi }$ of norm $%
\left
\Vert \varphi _{\chi }\right \Vert =1$, for every $\chi \in \widehat{L%
} $ which appears in the decomposition, we obtain an orthonormal basis which
we denote by $B_{L}$.

Since there exist $p+1$ different lines in $V$, we obtain in this manner a
collection of $p+1$ orthonormal bases, overall constructing a dictionary of
vectors $\mathfrak{D}_{H}=\left \{ \varphi \in B_{L}:L\subset V\right \} $
consisting of $p\left( p+1\right) $ vectors. We will call this dictionary,
for obvious reasons, the \textit{Heisenberg dictionary}.

The main property of the Heisenberg dictionary is summarized in the
following theorem \cite{H, HCM}

\begin{theorem}
\label{HeisCross_thm}For every pair of different lines $L,M\subset V$ and
for every $\varphi \in B_{L}$, $\phi \in B_{M}$ 
\begin{equation*}
\left \vert \left \langle \varphi ,\phi \right \rangle \right \vert =\frac{1%
}{\sqrt{p}}\text{.}
\end{equation*}
\end{theorem}

\subsubsection{The standard bases}

There are two standard examples of bases of the form $B_{L}$ associated with
the standard lines $T=\left \{ \left( \tau ,0\right) :\tau \in \mathbb{F}%
_{p}\right \} $ and $W=\left \{ \left( 0,w\right) :w\in \mathbb{F}%
_{p}\right
\} $. The basis $B_{W}$ consists of delta functions $\delta _{a}$%
, $a\in \mathbb{F}_{p},$ i.e., $\delta _{a}(t)=1$ if $a=t$ and $\delta
_{a}(t)=0$ otherwise, and the basis $B_{T}$ consists of normalized
characters $\psi _{a} $, $a\in \mathbb{F}_{p}$, where $\psi _{a}\left(
t\right) =1/\sqrt{p}\psi \left( at\right) $.

Indeed, the delta functions are common eigenfunctions of the operators $\pi
\left( 0,w\right) $, $w\in $\ $\mathbb{F}_{p}$ and the characters are common
eigenfunctions of the operators $\pi \left( \tau ,0\right) $, $\tau \in 
\mathbb{F}_{p}$.

Finally, for this specific example, the assertion of Theorem \ref%
{HeisCross_thm} amounts to%
\begin{equation}
|\left \langle \delta _{a},\psi _{b}\right \rangle |=\frac{1}{\sqrt{p}},
\label{corr_eq}
\end{equation}%
for every $\delta _{a}\in B_{W}$ and $\psi _{b}\in B_{T}$.

Theorem \ref{HeisCross_thm} asserts that (\ref{corr_eq}) holds for the
larger collection $\mathfrak{D}_{H}$ of $p+1$ orthonormal bases.

\subsection{The oscillator dictionary}

Reflecting back on the Heisenberg dictionary we see that it consists of a
collection of orthonormal bases characterized in terms of commutative
families of unitary operators where each such family is associated with a
commutative subgroup in the Heisenberg group $H$, via the Heisenberg
representation $\pi :H\rightarrow U\left( \mathcal{H}\right) $. In
comparison, the oscillator dictionary \cite{GHS} will be characterized in
terms of commutative families of unitary operators which are associated with
commutative subgroups in the symplectic group $Sp$ via the Weil
representation $\rho :Sp\rightarrow U\left( \mathcal{H}\right) $.

\subsubsection{Maximal tori\label{tori_sub}}

The commutative subgroups in $Sp$ that we consider are called maximal
algebraic tori \cite{B} (not to be confused with the notion of a topological
torus). A maximal (algebraic) torus in $Sp$ is a maximal commutative
subgroup which becomes diagonalizable over some field extension. The most
standard example of a maximal algebraic torus is the standard diagonal torus 
\begin{equation*}
A=\left \{ 
\begin{pmatrix}
a & 0 \\ 
0 & a^{-1}%
\end{pmatrix}%
:a\in \mathbb{F}_{p}^{\times }\right \} .
\end{equation*}

Standard linear algebra shows that up to conjugation\footnote{%
Two elements $h_{1},h_{2}$ in a group $G$ are called conjugated elements if
there exists an element $g\in G$ such that $g\cdot h_{1}\cdot g^{-1}=h_{2}$.
More generally, Two subgroups $H_{1},H_{2}\subset G$ are called conjugated
subgroups if there exists an element $g\in G$ such that $g\cdot H_{1}\cdot
g^{-1}=H_{2}$.} there exist two classes of maximal (algebraic) tori in $Sp$.
The first class consists of those tori which are diagonalizable already over 
$\mathbb{F}_{p}$, namely, those are tori $T$ which are conjugated to the
standard diagonal torus $A$ or more precisely such that there exists an
element $g\in Sp$ so that $g\cdot T\cdot g^{-1}=A$. A torus in this class is
called a \textit{split} torus.

The second class consists of those tori which become diagonalizable over the
quadratic extension $\mathbb{F}_{p^{2}}$, namely, those are tori which are
not conjugated to the standard diagonal torus $A$. A torus in this class is
called a \textit{non-split }torus (sometimes it is called inert torus).

All split (non-split) tori are conjugated to one another, therefore the
number of split tori is the number of elements in the coset space $Sp/N$
(see \cite{A} for basics of group theory), where $N$ is the normalizer group
of $A$; we have 
\begin{equation*}
\# \left( Sp/N\right) =\frac{p\left( p+1\right) }{2},
\end{equation*}%
and the number of non-split tori is the number of elements in the coset
space $Sp/M$, where $M$ is the normalizer group of some non-split torus; we
have%
\begin{equation*}
\# \left( Sp/M\right) =p\left( p-1\right) .
\end{equation*}

\paragraph{\textit{Example of a non-split maximal torus}}

It might be suggestive to explain further the notion of non-split torus by
exploring, first, the analogue notion in the more familiar setting of the
field $%
\mathbb{R}
$. Here, the standard example of a maximal non-split torus is the circle
group $SO(2)\subset SL_{2}(%
\mathbb{R}
)$. Indeed, it is a maximal commutative subgroup which becomes
diagonalizable when considered over the extension field $%
\mathbb{C}
$ of complex numbers. The above analogy suggests a way to construct examples
of maximal non-split tori in the finite field setting as well.

Let us assume for simplicity that $-1$ does not admit a square root in $%
\mathbb{F}_{p}$ or equivalently that $p\equiv 1\func{mod}4$. The group $Sp$
acts naturally on the plane $V=\mathbb{F}_{p}\times \mathbb{F}_{p}$.
Consider the standard symmetric form $B$ on $V$ given by 
\begin{equation*}
B((\tau ,w),(\tau ^{\prime },w^{\prime }))=\tau \tau ^{\prime }+ww^{\prime }.
\end{equation*}

An example of maximal non-split torus is the subgroup $SO=SO\left(
V,B\right) \subset Sp$ consisting of all elements $g\in Sp$ preserving the
form $B$, namely $g\in SO$ if and only if $B(gu,gv)=B(u,v)$ for every $%
u,v\in V$. In coordinates, $SO$ consists of all matrices $A\in SL_{2}\left( 
\mathbb{F}_{p}\right) $ which satisfy $AA^{t}=I$. The reader might think of $%
SO$ as the\ "finite circle".

\subsubsection{Bases associated with maximal tori}

Restricting the Weil representation to a maximal torus $T\subset Sp$, one
obtains a representation of a commutative group $\rho :T\rightarrow U\left( 
\mathcal{H}\right) $, which, in turns, yields an orthogonal decomposition
into character spaces (see Subsection \ref{char_sub}) 
\begin{equation}
\mathcal{H=}\tbigoplus_{\chi }\mathcal{H}_{\chi },  \label{decomp_eq}
\end{equation}%
where $\chi $ runs in the set $\widehat{T}$ of unitary characters of the
torus $T$, that is, each $\chi $ is a function $\chi :T\rightarrow S^{1},$
satisfying $\chi \left( t_{1}\cdot t_{2}\right) =\chi \left( t_{1}\right)
\cdot \chi \left( t_{2}\right) $, for every $t_{1},t_{2}\in T$.

A more concrete way to specify the above decomposition is by choosing a
generator\footnote{%
A maximal torus $T$ in $SL_{2}\left( \mathbb{F}_{p}\right) $ is a cyclic
group, thus there exists a generator.} $t_{0}\in T$, that is, an element
such that every $t\in T$ can be written in the form $t=t_{0}^{n}$, for some $%
n\in 
\mathbb{N}
$. After such a choice, the character space $\mathcal{H}_{\chi }$, which
appears in (\ref{decomp_eq}), naturally corresponds to the eigenspace of the
linear operator $\rho \left( t_{0}\right) $ associated to the eigenvalue $%
\lambda =\chi \left( t_{0}\right) $.

The decomposition (\ref{decomp_eq}) depends on the type of $T$:

\begin{itemize}
\item In the case where $T$ \ is a split torus we have $\dim \mathcal{H}%
_{\chi }=1$ unless $\chi =\sigma $, where $\sigma :T\rightarrow \left \{ \pm
1\right \} $ is the unique non-trivial quadratic character of $T$ (also
called the \textit{Legendre} character of $T$), in the latter case $\dim 
\mathcal{H}_{\sigma }=2$.

\item In the case where $T$ is a non-split torus then $\dim \mathcal{H}%
_{\chi }=1$ for every character $\chi $ which appears in the decomposition,
in this case the quadratic character $\sigma $ does not appear in the
decomposition (for details see \cite{GH}).
\end{itemize}

Choosing for every character $\chi \in \widehat{T},$ $\chi \neq \sigma $, a
vector $\varphi _{\chi }\in \mathcal{H}_{\chi }$ of unit norm, we obtain an
orthonormal system of vectors $B_{T}=\left \{ \varphi _{\chi }:\chi \neq
\sigma \right \} $, noting that in the case when $T$ is a non-split torus,
the set $B_{T}$ is, in fact, an orthonormal basis. Considering the union of
all these systems, we obtain the oscillator dictionary

\begin{equation*}
\mathfrak{D}_{O}=\left \{ \varphi \in B_{T}:T\subset Sp\right \} .
\end{equation*}

It is convenient to separate the dictionary $\mathfrak{D}_{O}$ into two
sub-dictionaries $\mathfrak{D}_{O}^{s}$ and $\mathfrak{D}_{O}^{ns}$ which
correspond to the split tori and the non-split tori respectively. The split
sub-dictionary $\mathfrak{D}_{O}^{s}$ consists of the union of all
orthonormal systems $B_{T}$, where $T$ runs through all the split tori in $%
Sp $, altogether $\tfrac{p(p+1)}{2}$ such systems, each consisting of $p-2$
orthonormal vectors, hence 
\begin{equation*}
\# \mathfrak{D}_{O}^{s}=\frac{p(p+1)(p-2)}{2}.
\end{equation*}

The non-split sub-dictionary $\mathfrak{D}_{O}^{ns}$ consists of the union
of all orthonormal bases $B_{T}$, where $T$ runs through all the non-split
tori in $Sp$, altogether $p(p-1)$ such bases, each consisting of $p$
orthonormal vectors, hence 
\begin{equation*}
\# \mathfrak{D}_{O}^{ns}=p^{2}(p-1).
\end{equation*}

Vectors in the oscillator dictionary satisfy many desired properties \cite%
{GHS}. In this paper we are only interested in the following property:

\begin{theorem}[\protect \cite{GHS}]
\label{OscCross_thm}Let $\phi \in B_{T_{1}}$ and $\varphi \in B_{T_{2}}$%
\begin{equation*}
\left \vert \left \langle \phi ,\varphi \right \rangle \right \vert \leq 
\frac{4}{\sqrt{p}}.
\end{equation*}
\end{theorem}

\paragraph{\textit{The system associated with the standard torus}}

It would be beneficial to give an explicit description of the system $B_{A}$
where $A\subset Sp$ is the standard diagonal torus, which is isomorphic to
the multiplicative group $G_{m}=\mathbb{F}_{p}^{\times }$. The torus $A$
acts on the Hilbert space $\mathcal{H}$ via the Weil representation yielding
a decomposition into character spaces 
\begin{equation*}
\mathcal{H=}\tbigoplus_{\chi \in \widehat{A}}\mathcal{H}_{\chi }.
\end{equation*}

For every $\chi \neq \sigma $ the character space $\mathcal{H}_{\chi }$ is
one dimensional. Our goal is to describe an explicit vector $\varphi _{\chi
}\in \mathcal{H}_{\chi }$ of unit norm: Let $\chi :G_{m}\rightarrow S^{1}$
be a non-trivial ($\chi \neq 1$) unitary character of the multiplicative
group. Thinking of the multiplicative group $G_{m}=\mathbb{F}_{p}^{\times }$
as sitting inside the line $\mathbb{F}_{p}$ we define the function $\varphi
_{\chi }\in \mathcal{%
\mathbb{C}
}\left( \mathbb{F}_{p}\right) $ as follows: 
\begin{equation*}
\varphi _{\chi }(t)=\left \{ 
\begin{array}{cc}
\frac{1}{\sqrt{p-1}}\chi (t) & t\neq 0 \\ 
0 & t=0%
\end{array}%
\right.
\end{equation*}%
Since, for every $a\in A$, $\rho \left( a\right) $ acts by normalized
scaling (see Subsection \ref{Wrep_sub}), it is easy to verify that $\varphi
_{\chi }$ is a character vector with respect to the action $\rho
:A\rightarrow U\left( \mathcal{H}\right) $ associated to the character $\chi
\cdot \sigma $.

Concluding, the orthonormal system $B_{A}$ \ is the set $\{ \varphi _{\chi
}:\chi \in \widehat{G}_{m},$ $\chi \neq 1\}$.

\subsubsection{Extended oscillator dictionary}

The oscillator dictionary can be extended to a much larger dictionary $%
\mathfrak{D}_{E}$ using the action of the Heisenberg group. Given a vector $%
\varphi \in \mathfrak{D}_{O}$ one can consider its orbit under the action of
the set $V\subset H$%
\begin{equation*}
\mathcal{O}_{\varphi }=V\cdot \varphi \triangleq \left \{ \pi \left(
v\right) \varphi :v\in V\right \} \text{.}
\end{equation*}

It is not hard to show that orbits associated to different vectors are
disjoint, therefore, we obtain a dictionary 
\begin{equation*}
\mathfrak{D}_{E}=\tbigcup \limits_{\varphi \in \mathfrak{D}}\mathcal{O}%
_{\varphi },
\end{equation*}%
consisting of $\# \left( V\right) \cdot \# \left( \mathfrak{D}_{O}\right)
\sim p^{5}$ vectors. Interestingly, the extended dictionary $\mathfrak{D}%
_{E} $ continues to be $4/\sqrt{p}$ - coherent, this is a consequence of the
following generalization of Theorem \ref{OscCross_thm}:

\begin{theorem}[\protect \cite{GH}]
\label{stability_thm}Given two vectors $\varphi ,\phi \in \mathfrak{D}_{O}$
and an element $0\neq v\in V$ we have 
\begin{equation*}
\left \vert \left \langle \varphi ,\pi \left( v\right) \phi \right \rangle
\right \vert \leq \frac{4}{\sqrt{p}}\text{.}
\end{equation*}
\end{theorem}

For a proof, see \cite{GHS}.

\begin{remark}
A way to interpret Theorem \ref{stability_thm} is to say that any two
different vectors $\varphi \neq \phi \in \mathfrak{D}_{O}$ are incoherent in
a stable sense, that is, their coherency is $4/\sqrt{p}$ no matter if any
one of them undergoes an arbitrary time/phase shift. This property seems to
be important in communication where a transmitted signal may acquire time
shift due to asynchronous communication and phase shift due to Doppler
effect.
\end{remark}

\bigskip \appendix

\section{Terminology from representation theory\label{TGR_sec}}

\subsection{Finite fields}

Given a prime number $p\geq 2$, there exists a unique finite field
consisting of $p$ elements, denoted by $\mathbb{F}_{p}$. A way to visualize
this field is as a discrete set of $p$ points, cyclically ordered and
indexed by the numbers $0,1,...,p-1$.

Most of the constructions of linear algebra carry over to the finite field
setting, in particular one can consider matrix groups with matrix entries
from $\mathbb{F}_{p}$. Particular examples of such groups which play a role
in this paper are the special linear group $SL_{2}\left( \mathbb{F}%
_{p}\right) $ and the special orthogonal group $SO_{2}\left( \mathbb{F}%
_{p}\right) $. The first, consists of $2\times 2$ matrices 
\begin{equation*}
\begin{pmatrix}
a & b \\ 
c & d%
\end{pmatrix}%
,\text{ }a,b,c,d\in \mathbb{F}_{p},
\end{equation*}%
such that $ad-bc=1$. The second is a subgroup of $SL_{2}\left( \mathbb{F}%
_{p}\right) $ consisting of matrices $A\in SL_{2}\left( \mathbb{F}%
_{p}\right) $ such that $AA^{t}=Id$.

\subsection{Unitary representations\label{rep_sub}}

Let $\mathcal{H}$ be a finite dimensional complex Hilbert space, equipped
with an inner product $\left \langle \cdot ,\cdot \right \rangle :\mathcal{%
H\times H\rightarrow 
\mathbb{C}
}$. A unitary operator on $\mathcal{H}$ is an operator $A:\mathcal{%
H\rightarrow H}$ which preserves the inner product, that is, $\left \langle
Af,Ag\right \rangle =\left \langle f,g\right \rangle $, for every $f,g\in 
\mathcal{H}$. The set of unitary operators forms a group under composition
of operators, which is denoted by $U\left( \mathcal{H}\right) $.

We proceed to introduce the notion of a unitary representation (see \cite{A,
Se} for a more comprehensive treatment). Let $G$ be a finite group.

\begin{definition}
A unitary representation of $G$ on the Hilbert space $\mathcal{H}$ is a
homomorphism $\pi :G\rightarrow U\left( \mathcal{H}\right) $, that is, $\pi $
is a map which satisfies the condition 
\begin{equation*}
\pi \left( g\cdot h\right) =\pi \left( g\right) \circ \pi \left( h\right) ,
\end{equation*}%
for every $g,h\in G$.
\end{definition}

Specifying a unitary representation $\pi :G\rightarrow U\left( \mathcal{H}%
\right) $ gives a convenient way to think of the collection of unitary
operators $\left \{ \pi \left( g\right) :g\in G\right \} $ and the relations
that they satisfy between one another - these relations are encoded in the
structure of the group $G$.

The smallest unitary representations are the\textit{\ irreducible} unitary
representations.

\begin{definition}
A unitary representation $\pi :G\rightarrow U\left( \mathcal{H}\right) 
\mathcal{\ }$is called irreducible if there is no proper vector space $0\neq 
\mathcal{H}^{\prime }\subsetneqq \mathcal{H}$ invariant under $G$, i.e.,
such that 
\begin{equation*}
\pi (g)[f]\in \mathcal{H}^{\prime },
\end{equation*}%
for every $f\in $ $\mathcal{H}^{\prime }$.
\end{definition}

Irreducible unitary representations form the building blocks of all unitary
representations in the sense that every unitary representation $\pi
:G\rightarrow U\left( \mathcal{H}\right) $ can be decomposed into a direct
sum of irreducible unitary representations. The precise statement is that
always there exists is a decomposition of the Hilbert space $\mathcal{H}$
into a direct sum 
\begin{equation*}
\mathcal{H=}\tbigoplus \limits_{i\in I}\mathcal{H}_{i}\text{,}
\end{equation*}%
such that each subspace $\mathcal{H}_{i}$ is closed under the action of $G$,
that is $\pi \left( g\right) \left[ f\right] \in \mathcal{H}_{i}$, for every 
$f\in \mathcal{H}_{i}$ and such that the restricted unitary representations $%
\pi _{i}:G\rightarrow U\left( \mathcal{H}_{i}\right) $ are irreducible.

\subsubsection{Unitary representations of commutative groups}

A particular situation occurs when $G$ is a commutative group, that is, a
group for which $g\cdot h=h\cdot g$, for every $g,h\in G$. In this
situation, specifying a unitary representation $\pi :G\rightarrow U\left( 
\mathcal{H}\right) $ is equivalent to specifying a collection of unitary
operators $\left \{ \pi \left( g\right) :g\in G\right \} $ which commute
pairwisely; this follows from the following relation: 
\begin{equation*}
\pi \left( g\right) \circ \pi \left( h\right) =\pi \left( g\cdot h\right)
=\pi \left( h\cdot g\right) =\pi \left( h\right) \circ \pi \left( g\right) ,
\end{equation*}%
for every $g,h\in G$.

\subsection{Basic examples\label{examples_sub}}

We proceed to describe three basic examples of unitary representations of
commutative groups, which are of particular relevance to this paper. In all
these examples the Hilbert space is taken to be $\mathcal{H=%
\mathbb{C}
}\left( \mathbb{F}_{p}\right) $.

\subsubsection{Time shifts}

Let $\left( \mathbb{F}_{p},+\right) $ be the additive group, let us denote
the parameter of $\mathbb{F}_{p}$ by $\tau $. $\ $Define the unitary
representation $L:\mathbb{F}_{p}\rightarrow U\left( \mathcal{H}\right) $
given by $\tau \mapsto L_{\tau }$, where $L_{\tau }$ is the unitary operator
of cyclic time translation by $\tau $:%
\begin{equation*}
L_{\tau }\left[ f\right] \left( t\right) =f\left( t+\tau \right) ,
\end{equation*}%
for every $f\in \mathcal{H}$.

\subsubsection{Phase shifts}

Let $\left( \mathbb{F}_{p},+\right) $ be the same as in the previous
example, let us denote the parameter of $\mathbb{F}_{p}$ by $w$. Define the
unitary representation $M:\mathbb{F}_{p}\rightarrow U\left( \mathcal{H}%
\right) $ given by $w\mapsto M_{w}$, where $M_{w}$ is the unitary operator
of cyclic phase translation by $w$:%
\begin{equation*}
M_{w}\left[ f\right] \left( t\right) =e^{\frac{2\pi i}{p}wt}f\left( t\right)
,
\end{equation*}%
for every $f\in \mathcal{H}$.

\subsubsection{Scaling}

Let $\left( \mathbb{F}_{p}^{\times },\cdot \right) $ be the multiplicative
group, let us denote the parameter of $\mathbb{F}_{p}^{\times }$ by $a$.
Define the unitary representation $S:\mathbb{F}_{p}^{\times }\rightarrow
U\left( \mathcal{H}\right) $ given by $a\mapsto S_{a}$, where $S_{a}$ is the
unitary operator\ of scaling by $a$: 
\begin{equation*}
S_{a}\left[ f\right] \left( t\right) =f\left( a\cdot t\right) ,
\end{equation*}%
for every $f\in \mathcal{H}$.

\subsection{Intertwining morphisms \label{inter_sub}}

Let $\pi _{i}:G\rightarrow U\left( \mathcal{H}_{i}\right) $, $i=1,2$, be a
pair of unitary representations.

\begin{definition}
An intertwining morphism from $\pi _{1}$ to $\pi _{2}$ is a unitary operator 
$A:\mathcal{H}_{1}\rightarrow \mathcal{H}_{2}$ which satisfies 
\begin{equation*}
A\circ \pi _{1}\left( g\right) [f]=\pi _{2}\left( g\right) \circ A[f],
\end{equation*}%
for every $g\in G$ and for every $f\in \mathcal{H}_{1}.$
\end{definition}

The space of intertwining morphisms from $\pi _{1}$ to $\pi _{2}$ is denoted
by $\mathrm{Hom}_{G}\left( \pi _{1},\pi _{2}\right) $. In addition, if there
exist $A\in \mathrm{Hom}_{G}\left( \pi _{1},\pi _{2}\right) $ which is also
a bijection then we say that $\pi _{1}$ and $\pi _{2}$ are \textit{%
isomorphic }unitary\textit{\ }representations. An elementary but useful
result is the so called Schur's lemma:

\begin{lemma}[Schur's Lemma]
Let $\pi :G\rightarrow U\left( \mathcal{H}\right) $ be an irreducible
unitary representation then every intertwining morphism $A\in \mathrm{Hom}%
_{G}\left( \pi ,\pi \right) $ is a scalar operator, i.e., $A=a\cdot Id_{%
\mathcal{H}}$ for some $a\in S^{1}$.
\end{lemma}

For a proof, see \cite{A, Se}.

\subsection{Character vectors\label{char_sub}}

The final piece of terminology that we will require is the notion of a
character vector, which generalizes the notion of an eigenvector.

First, recall the following basic fact from linear algebra:

\textbf{Fact 1. } A unitary operator $A:\mathcal{H\rightarrow H}$ can be
diagonalized, which means that there exists an orthogonal decomposition of $%
\mathcal{H}$ into a direct sum of eigenspaces 
\begin{equation*}
\mathcal{H=}\tbigoplus \limits_{\lambda \in S^{1}}\mathcal{H}_{\lambda },
\end{equation*}%
where for $\varphi \in \mathcal{H}_{\lambda }$ we have $A\varphi =\lambda
\varphi $.

The more general situation occurs when one consider a unitary representation 
$\pi :G\rightarrow U\left( \mathcal{H}\right) $ of a commutative group $G$.
Such a representation yields a collection $\left \{ \pi \left( g\right)
:g\in G\right \} $ of unitary operators which commute pairwisely.

\textbf{Fact 2. }The unitary operators $\left \{ \pi \left( g\right) :g\in
G\right \} $ can be diagonalized simultaneously, which means that there
exists an orthogonal decomposition of $\mathcal{H}$ into common eigenspaces%
\begin{equation*}
\mathcal{H=}\tbigoplus \limits_{\chi :G\rightarrow S^{1}}\mathcal{H}_{\chi }.
\end{equation*}

The common eigenspaces are now indexed by functions $\chi :G\rightarrow
S^{1} $ where each function $\chi $ encodes the eigenvalues associated with
the different operators $\pi \left( g\right) $, $g\in $ $G$. In more
details, for $\varphi \in \mathcal{H}_{\chi }$, we have $\pi \left( g\right)
\varphi =\chi \left( g\right) \varphi $, for every $g\in G$.\ 

It is easy to verify that the functions $\chi $ which appear in the above
decomposition are unitary characters of the group $G$, that is, $\chi \left(
g\cdot h\right) =\chi \left( g\right) \cdot \chi \left( h\right) $, for
every $g,h\in G$.

The spaces $\mathcal{H}_{\chi }$ are called a \textit{character spaces} and
a vector $\varphi \in \mathcal{H}_{\chi }$ is called \textit{character vector%
}.

\subsection{Basic decompositions}

Considering our three basic examples (see Subsection \ref{examples_sub}), we
obtain, respectively, three orthogonal decompositions of $\mathcal{H}$ into
character spaces.

\subsubsection{Time shift invariant decomposition}

For every $w\in \mathbb{F}_{p}$, let $\psi _{w}:\mathbb{F}_{p}\rightarrow
S^{1}$ denote the character $\psi _{w}\left( \tau \right) =e^{\frac{2\pi i}{p%
}w\tau }$. The decomposition into character spaces with respect to the
representation $L$ is 
\begin{equation*}
\mathcal{H=}\tbigoplus \limits_{w\in \mathbb{F}_{p}}\mathcal{H}_{\psi _{w}}%
\text{,}
\end{equation*}%
with $\dim $ $\mathcal{H}_{\psi _{w}}=1$, for every $w\in \mathbb{F}_{p}$. A
function $\varphi \in \mathcal{H}_{\psi _{w}}$ if $\varphi \left( t\right)
=c\cdot e^{\frac{2\pi i}{p}wt}$, for some constant $c\in 
\mathbb{C}
$.

\subsubsection{Phase shift invariant decomposition}

For every $\tau \in \mathbb{F}_{p}$, let $\psi _{\tau }:\mathbb{F}%
_{p}\rightarrow S^{1}$ denote the character $\psi _{\tau }\left( w\right)
=e^{\frac{2\pi i}{p}\tau w}$. The decomposition into character spaces with
respect to the representation $M$ is 
\begin{equation*}
\mathcal{H=}\tbigoplus \limits_{\tau \in \mathbb{F}_{p}}\mathcal{H}_{\psi
_{\tau }}\text{,}
\end{equation*}%
with $\dim $ $\mathcal{H}_{\psi _{\tau }}=1$, for every $\tau \in \mathbb{F}%
_{p}$. A function $\varphi \in \mathcal{H}_{\psi _{\tau }}$ if $\varphi
=c\cdot \delta _{\tau }$, for some constant $c\in 
\mathbb{C}
$ and 
\begin{equation*}
\delta _{\tau }\left( t\right) =\left \{ 
\begin{array}{cc}
1, & t=\tau , \\ 
0, & t\neq \tau .%
\end{array}%
\right.
\end{equation*}

\subsubsection{Scale invariant decomposition}

let us first explain how to describe unitary characters of the
multiplicative group $\left( \mathbb{F}_{p}^{\times },\cdot \right) $. The
basic fact that we use is that $\mathbb{F}_{p}^{\times }$ is a cyclic group
of order $p-1$, which means that we can write $\mathbb{F}_{p}^{\times }$ in
the form $\left \{ 1,r,...,r^{p-1}\right \} $, for some generator $r\in 
\mathbb{F}_{p}^{\times }$. This implies that unitary characters can be
specified as follows: For a $\left( p-1\right) $th roots of unity, $\zeta
\in \mu _{p-1}$, let $\chi _{\zeta }:\mathbb{F}_{p}^{\times }\rightarrow
S^{1}$ be the unitary character given by 
\begin{equation*}
\chi _{\zeta }\left( r^{k}\right) =\zeta ^{k}.
\end{equation*}

Now, the decomposition into character spaces with respect to the
representation $S$ is 
\begin{equation*}
\mathcal{H=}\tbigoplus \limits_{\zeta \in \mathbb{\mu }_{p-1}}\mathcal{H}%
_{\chi _{\zeta }}\text{,}
\end{equation*}%
with $\dim $ $\mathcal{H}_{\chi _{\zeta }}=1$, for every $\zeta \neq 1$ and $%
\dim $ $\mathcal{H}_{\chi _{\zeta }}=2$, for $\zeta =1$. The one dimensional
space $\mathcal{H}_{\chi _{\zeta }}$, $\zeta \neq 1$ is spanned by the
function 
\begin{equation*}
\varphi _{\zeta }\left( t\right) =\left \{ 
\begin{array}{cc}
\chi _{\zeta }\left( t\right) , & t\neq 0, \\ 
0, & t=0,%
\end{array}%
\right.
\end{equation*}%
and the two dimensional space $\mathcal{H}_{\chi _{\zeta }}$, $\zeta =1$ is
spanned by the function $\delta _{0}$ and the constant function \thinspace $%
1 $.

\section{\textbf{Construction of the oscillator dictionary\label{Algorithm}}}

\subsection{Algorithm}

We describe an explicit\ algorithm that generates the oscillator dictionary $%
\mathfrak{D}_{O}^{s}$ associated with the collection of split tori in $Sp$.

\subsubsection{Tori}

Consider the standard diagonal torus 
\begin{equation*}
A=\left \{ 
\begin{pmatrix}
a & 0 \\ 
0 & a^{-1}%
\end{pmatrix}%
:\text{ }a\in \mathbb{F}_{p}^{\times }\right \} .
\end{equation*}

Every split torus in $Sp$ is conjugated to the torus $A$, which means that
the collection $\mathcal{T}$ of all split tori in $Sp$ can be written as 
\begin{equation*}
\mathcal{T}=\{gAg^{-1}:\ g\in Sp\}.
\end{equation*}

\subsubsection{Parametrization}

A direct calculation reveals that every torus in $\mathcal{T}$ can be
written as $gAg^{-1}$ for an element $g$ of the form 
\begin{equation}
g=%
\begin{pmatrix}
1 & b \\ 
c & 1+bc%
\end{pmatrix}%
,\text{ }b,c\in \mathbb{F}_{p}.  \label{form_eq}
\end{equation}

If $b=0$, this presentation is unique: In the case $b\neq 0$, an element $%
\widetilde{g}$ represents the same torus as $g$ if and only if \ it is of
the form 
\begin{equation*}
\widetilde{g}=%
\begin{pmatrix}
1 & b \\ 
c & 1+bc%
\end{pmatrix}%
\begin{pmatrix}
0 & -b \\ 
b^{-1} & 0%
\end{pmatrix}%
.
\end{equation*}

Let us choose a set of elements of the form (\ref{form_eq}) representing
each torus in $\mathcal{T}$ exactly once and denote this set of
representative elements by $R$. $.$

\subsubsection{Generators}

The group $A$ is a cyclic group and we can find a generator $g_{A}$ for $A$.
This task is simple from the computational perspective, since the group $A$
is finite, consisting of $p-1$ elements.

Now, we make the following two observations. First observation is that the
oscillator basis $B_{A}$ is the basis of eigenfunctions of the operator $%
\rho \left( g_{A}\right) $.

The second observation is, that other bases in the oscillator system $%
\mathfrak{D}_{O}^{s}$ can be obtained from $B_{A}$ by applying elements from
the set $R$. More specifically, for a torus $T$ of the form $T=gAg^{-1}$, $%
g\in R,$ we have 
\begin{equation*}
B_{gAg^{-1}}=\{ \rho (g)\varphi :\text{ }\varphi \in B_{A}\}.
\end{equation*}

Concluding, we described the (split) oscillator system as%
\begin{equation*}
\mathfrak{D}_{O}^{s}\mathcal{=\{}\rho \left( g\right) \varphi :g\in
R,\varphi \in B_{A}\}.
\end{equation*}

\subsubsection{Formulas}

We are left to explain how to write explicit formulas (matrices) for the
operators $\rho \left( g\right) $, $g\in R$.

First, we recall that the group $Sp$ admits a Bruhat decomposition $Sp=B\cup
B\mathrm{w}B,$ where $B$ is the Borel subgroup consisting of lower
triangular matrices in $Sp$ and $\mathrm{w}$ denotes the Weyl element%
\begin{equation*}
\mathrm{w}=%
\begin{pmatrix}
{\small 0} & {\small 1} \\ 
-{\small 1} & {\small 0}%
\end{pmatrix}%
.
\end{equation*}

Furthermore, the Borel subgroup $B$ can be written as a product $B=AU=UA$,
where $A$ is the standard diagonal torus and $U$ is the standard unipotent
group 
\begin{equation*}
U=\left \{ 
\begin{pmatrix}
1 & 0 \\ 
u & 1%
\end{pmatrix}%
:u\in \mathbb{F}_{p}\right \} ,
\end{equation*}%
therefore, we can write the Bruhat decomposition also as $Sp=UA\cup UA%
\mathrm{w}U$.

Using the Bruhat decomposition we conclude that every operator $\rho \left(
g\right) $, $g\in Sp$, can be written either in the form $\rho \left(
g\right) =M_{u}\circ S_{a}$ or in the form $\rho \left( g\right)
=M_{u_{2}}\circ S_{a}\circ F\circ M_{u_{1}}$, where $M_{u},S_{a}$ and $F$
are the explicit operators which appears in the description of the Weil
representation in Subsection \ref{Wrep_sub}.

\begin{example}
For $g\in R$, with $b\neq 0$, the Bruhat decomposition of $g$ is given
explicitly by 
\begin{equation*}
g=%
\begin{pmatrix}
1 & 0 \\ 
\frac{1+bc}{b} & 1%
\end{pmatrix}%
\begin{pmatrix}
b & 0 \\ 
0 & b^{-1}%
\end{pmatrix}%
\begin{pmatrix}
0 & 1 \\ 
-1 & 0%
\end{pmatrix}%
\begin{pmatrix}
1 & 0 \\ 
b^{-1} & 1%
\end{pmatrix}%
,
\end{equation*}%
and consequently 
\begin{equation*}
\rho \left( g\right) =M_{\frac{1+bc}{b}}\circ S_{b}\circ F\circ M_{b^{-1}}.
\end{equation*}%
For $g\in R$, with $c=0,$ we have%
\begin{equation*}
g=%
\begin{pmatrix}
1 & 0 \\ 
u & 1%
\end{pmatrix}%
,
\end{equation*}%
and 
\begin{equation*}
\rho \left( g\right) =M_{u}.
\end{equation*}
\end{example}

\subsection{Pseudocode}

Below, is given a pseudo-code description of the construction of the \textit{%
oscillator dictionary }$\mathfrak{D}_{O}^{s}.$

\begin{enumerate}
\item Choose a prime\textrm{\ }$p.$

\item Compute generator\textrm{\ }$g_{A}$ for the standard torus $A$.

\item Diagonalize\textrm{\ }$\rho \left( g_{A}\right) $ and obtain the basis
of eigenfunctions $\mathcal{B}_{A}.$

\item For every\textrm{\ }$g\in R$:

\item Compute the operator\textrm{\ }$\rho \left( g\right) $ as follows:

\begin{enumerate}
\item Calculate the Bruhat decomposition of\textrm{\ }$g$, namely, write $g$
in the form $g=u_{2}\cdot a\cdot \mathrm{w}\cdot u_{1}$ or $g=u\cdot a$.

\item Calculate the operator\textrm{\ }$\rho \left( g\right) $, namely, take 
$\rho \left( g\right) =M_{u_{2}}\circ S_{a}\circ F\circ M_{u_{1}}$ or $\rho
\left( g\right) =M_{u}\circ S_{a}$.
\end{enumerate}

\item Compute the vectors $\rho (g)\varphi $, for every $\varphi \in B_{A}\ $%
and obtain the system $B_{gAg^{-1}}$.
\end{enumerate}

\begin{remark}[Running time]
It is easy to verify that the time complexity of the algorithm presented
above is $O(p^{4}\log p)$. This is, in fact, an optimal time complexity,
since already to specify $p^{3}$ vectors, each of length $p$, requires $%
p^{4} $ operations.
\end{remark}

\end{document}